\documentclass[twocolumn,nopacs, preprintnumbers, amsmath, amssymb, superscriptaddress,pra]{revtex4}
\usepackage{graphicx}
\usepackage{dcolumn}
\usepackage{bm}
\usepackage{epsf}
\usepackage{amssymb}
\DeclareGraphicsExtensions{.eps}
\usepackage{epstopdf}
\usepackage{natbib}
\usepackage{subfigure}
\usepackage{graphicx}
\begin{document}

\author{David S. Simon}
\affiliation{Dept. of Electrical and Computer Engineering, Boston
University, 8 Saint Mary's Street, Boston, MA 02215}
\affiliation{Dept. of Physics and Astronomy, Stonehill College, 320 Washington Street, Easton, MA 02357}


\title{Quantum Sensors: Improved Optical Measurement via Specialized Quantum States}


\maketitle


\section*{Abstract}\label{abstractsection}
Classical measurement strategies in many areas are approaching their maximum resolution and sensitivity levels, but these levels often still fall far short of the
ultimate limits allowed by the laws of physics. To go further, strategies must be adopted that take into account the quantum nature of the probe particles and that optimize their quantum states for the desired application. Here
we review some of these approaches, in which quantum entanglement, the orbital angular momentum of single photons, and quantum interferometry are used to produce
optical measurements beyond the classical limit.

\section{Introduction}\label{introsection}

In the search for greater sensitivity and resolution in measuring and sensing applications, there are fundamental limits imposed by the laws of physics. In order
to reach these limits, it has become clear that it is ultimately necessary to take into account the fundamentally quantum mechanical nature of both the system
being measured and of the devices that are used to measure them. The goal of this review is to give an introduction to some of the ideas needed in order to continue
pushing measurement capabilities of future devices toward their ultimate capacity.

In imaging applications, it has long been common lore that the Abb\'e limit prevents an optical system from resolving features smaller than the wavelength of the illuminating light.
However, it has been recognized in recent years that there are loopholes in the Abb\'e limit that allow so-called \emph{super-resolution}, the resolution of
sub-wavelength-sized features \cite{lukosz,heintzmann}. This can be done by using (i) spatially-structured light and (ii) nonlinear interactions between the
light and the object. The spatial structure means that the wavefronts have some nontrivial set of spatial frequencies, which in a sense combine with the light's
intrinsic wavelength to produce a smaller \emph{effective} wavelength, thus allowing the resolving of smaller features. To detect this structure, some nonlinear
interaction is utilized, such as the two-photon absorption processes commonly used in microscopy.

Similar "super-resolution" or "super-sensitivity" effects also occur in other measurement and sensing contexts outside of the realm of imaging, and these effects
are often associated with purely quantum mechanical phenomena such as entanglement. Entanglement occurs when a composite system, made of two or more subsystems,
is described by a single quantum state for the entire system, in such a way that it cannot be divided into separate well-defined substates for the individual
subsystems. (See the appendix for a more precise definition of entanglement.) In this review, we give a short description of some of these quantum-based
super-resolving effects in areas other than standard imaging. Specifically, we will focus on the use of non-classical optical states to measure quantities
related to phase, angular and rotational variables, and dispersive properties. We will see that in each of these cases the use of quantum-based methods allows
more precise measurements than is possible through purely classical means. For example, entangled pairs of photons can carry out phase measurements that violate
the so-called standard quantum limit (see the next section). In many of these cases, quantum entanglement can be seen as a resource that may be used to carry out
practical tasks, a viewpoint is common in the quantum computation and quantum information communities, but is not as widely recognized in other areas. The
improved resolution and sensitivity that can be obtained from an $N$-photon entangled state can be seen as being due to the fact that the set of photons (each of
wavelength $\lambda$) can be treated as a single object (essentially a small Bose-Einstein condensate) with a single joint de Broglie wavelength ${\lambda \over
\sqrt{N}}$ \cite{jacobson,fonseca,edamatsu,mitchell,walther,resche}. In this way, the true quantum-mechanical limit, the Heisenberg limit, can be be reached.

By optimizing the detection method for a given measurement, the weaker standard quantum limit can be achieved, but to reach the stronger Heisenberg limit, it is
necessary to optimize the probe state as well \cite{giov1} and tailor its  properties to the desired measurement; this is where entanglement generally enters the
picture. The comparison to super-resolution imaging also holds here: the Abb\'e limit can be reached by optimizing the detection system (large-aperture
lenses, high resolution detectors, etc.), but to exceed the limit it is necessary to also optimize the illuminating light (the probe state) by introducing
appropriate spatial structure. As in the imaging case, nonlinear processes are also involved in non-imaging quantum sensing applications as well: nonlinear
interactions in solids (spontaneous parametric down conversion \cite{shih} in nonlinear crystals) are commonly used to prepare the entangled probe state, and the
detection process itself typically involves coincidence counting or other forms of multiphoton detection: such methods are essentially nonlinear since the
signals from several photons are multiplied for each detection event. In particular, two-detector coincidence counting is often used, which measures the
second-order (intensity-intensity) correlations of a system, as opposed to the first-order (amplitude-amplitude) correlations measured in classical
interferometry. Many purely quantum mechanical effects that are invisible to classical interferometry show up in the second-order correlations
\cite{mandel,loudon}. In addition, interference effects in such measurements have a maximum visibility of ${1\over \sqrt{2}}\approx 71\%$ for classical states of
light \cite{clauser1,clauser2}, whereas entangled states can have visibilities approaching $100\%$.

A review of the more strictly technological aspects of quantum sensing, such as the development of on-chip photon sources and interferometers can be found in
\cite{obrien}, and various other aspects of quantum metrology have previously been reviewed in \cite{giov1,greggsasha,kapale,giov2}. Here we concentrate on the physical ideas and some specific applications. The main ingredients in all of these applications are (i) entanglement, (ii) using states tailored to optimize the given measurement, and (iii) coincidence counting or intensity-intensity correlations.
Although most of the applications in this area are still confined to the
lab, some have begun to make steps toward real-world use. As one example, some of the methods described here have recently been used to
measure the polarization mode dispersion of a switching unit in a fiber optics communication network \cite{fraine1,fraine2}; the resulting measurement had
uncertainties of $0.1$ fs for group delay and $2$ attoseconds for phase delay, which is orders of magnitude smaller than the uncertainties present in previous
classical measurement strategies.

One note on terminology: super-resolution is taken to mean producing interference patterns that oscillate on smaller scales than expected classically, while
super-sensitivity means reducing the uncertainty in a measured variable below its classical value. The two effects often go hand-in-hand.

A very brief summary of the
necessary quantum mechanical background is given in the appendix.

\section{Phase Estimation, Entanglement, and Quantum Limits}

We consider phase measurements first \cite{kapale,hradil2}, and ask what fundamental limit does quantum mechanics place on our ability to measure them? The question is muddled a bit by the following problem. In quantum mechanics, physical variables are represented by operators (see the appendix), and we would
expect the photon number operator $\hat N$ and the phase operator $\hat \phi$ to be canonically conjugate operators, with their commutator leading to an uncertainty relation that defines the ultimate physical limit on measurements. However, it has been known since at least the 1960's that it is impossible to define a
unitary operator $\widehat{e^{i\phi}}$, or equivalently a Hermitian operator $\hat \phi$, that gives the correct canonical commutation relations. There is a
large literature discussing this problem \cite{sussglo,carr,levy,pegg1,vaccarro1,barnett1,vaccarro2,barnett2,vacarro3,summy,barnett3,pegg2,bergouenglert}, and
there are several possible resolutions to it. Here we will simply follow the most common path and make use of the non-Hermitian Susskind-Glogower (SG) operator
\cite{sussglo}
\begin{equation}\hat S=\widehat{e^{i\phi}}=\sum_{n=0}^\infty |n\rangle \langle n+1| =\hat a\hat N^{-1/2} =(\hat N+1)^{-1/2} \hat a,\end{equation} where,
$\hat a$, $\hat N$, and $|n\rangle$ are, respectively, the photon annihilation operator, the photon number operator, and the Fock state of photon number $n$. The
eigenstates of $\hat S$ are $|\phi\rangle =\sum_{n=0}^\infty e^{in\phi}|n\rangle $. The corresponding eigenvalue relation is given by $\hat S|\phi\rangle =
e^{i\phi}|\phi\rangle$, where $\phi $ is the phase of the state (up to some unobservable overall reference phase). To get a Hermitian (and therefore physically
measurable) operator, we can add $\hat S$ to its Hermitian conjugate to form the new operator $\hat A= \sum_{n=0}^\infty \left( |n\rangle \langle n+1| +
|n+1\rangle \langle n|\right) .$ The expectation value of this operator measures the rate of single photon transitions within the apparatus. Now that we have a
phase-dependent observable, we can measure its value on some quantum state and extract an estimate of the state's phase. (Note that any other phase-dependent
Hermitian operator could be used instead; however, some operators give better phase estimation than others and are easier to measure experimentally.)

Suppose the phase is measured using an interferometer, with input to the system only at a single port. We can consider several types of input, such as a coherent state input
(where the number of photons fluctuates about some mean value $N$ according to Poisson statistics), a single Fock state (of fixed photon number $N$), or $N$ separate measurements on a stream of
single-photon states. In each of these cases, the phase sensitivity (the minimum uncertainty in the measured phase value) scales for large $N$ as ${1\over
\sqrt{N}}$. We can see this most easily by looking at the case of $N$ measurements of consecutive single photon inputs.

\begin{figure}
\centering
\includegraphics[scale=.3]{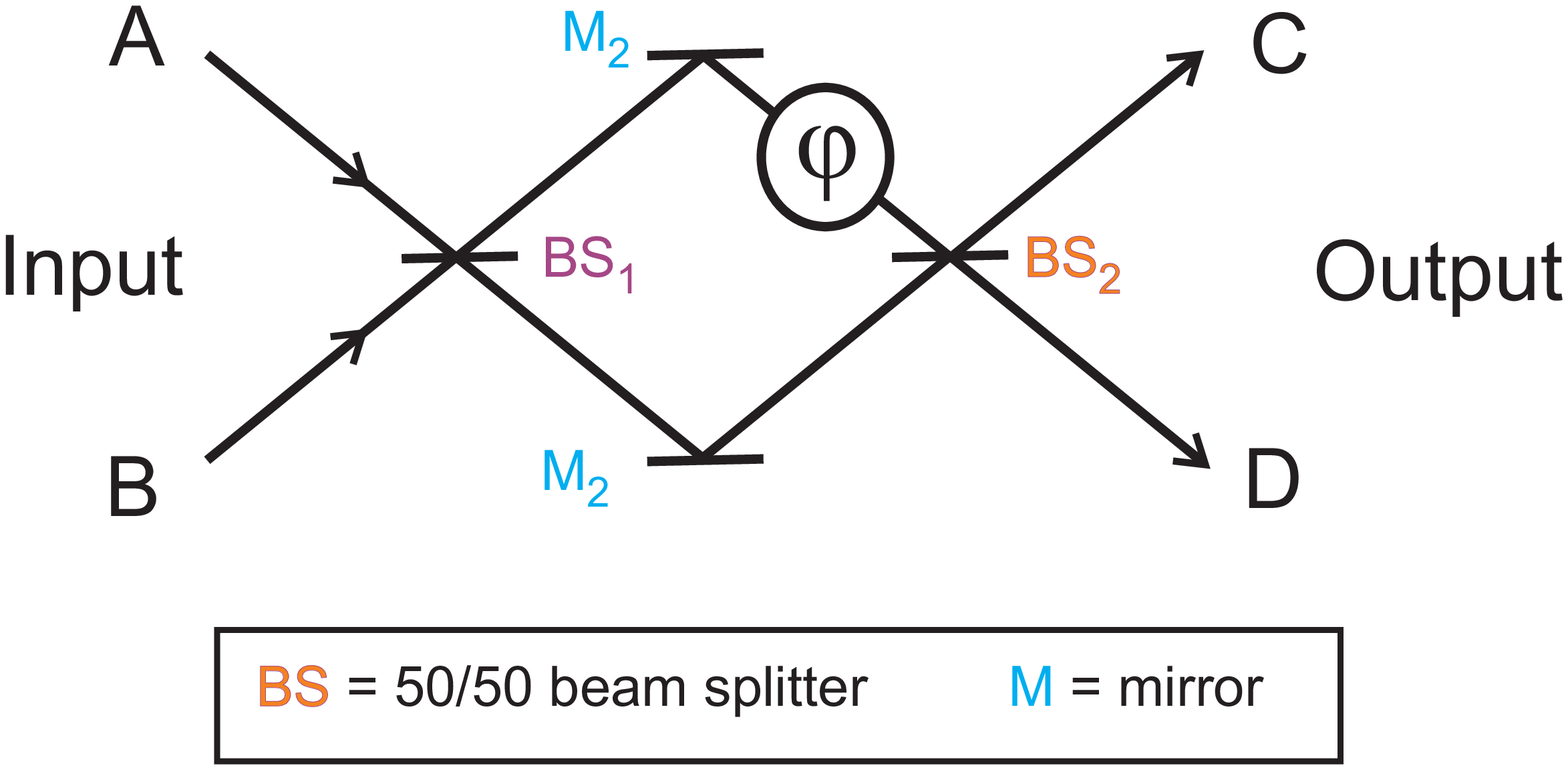}
\caption{A Mach-Zehnder interferometer. Input at either of the input ports ($A$ or $B$) has two possible ways of reaching the final beam splitter, $BS_2$.
Along the upper path, a phase shift of $\phi$ is added. The second beam splitter makes it impossible to determine which path was followed simply from looking at the output at ports $C$ and $D$. As a result, the amplitudes for the both possibilities must be present, leading to interference at the output ports.}
\label{mzfig}       
\end{figure}

Consider a photon entering port $A$ in the Mach-Zehnder interferometer of Figure \ref{mzfig}. It can travel along the lower path, reflecting off mirror $M_2$, or
it can travel along the upper path, reflecting off $M_1$ and gaining a phase shift of $\phi$. (We will neglect any phase shift due to reflection at the beam
splitter, since it can be absorbed into the definition of $\phi$.) We may denote the state leaving the first beam splitter along the lower path by $|L\rangle$
and the state leaving along the upper path by $|U\rangle$. (See the appendix for an explanation of state vector notation.) Or equivalently we could label these
two states by the number of photons in the \emph{upper} branch, so that $|L\rangle =|0\rangle$ and $|U\rangle =|1\rangle$. Then the state arriving at the second
beam splitter is $|\psi \rangle ={1\over \sqrt{2}}\left(|0\rangle +e^{i\phi } |1\rangle \right).$ Since the photon number $n$ never exceeds $1$ anywhere for this
state, the operator $\hat A$ can be truncated to $\hat A=|0\rangle \langle 1| +|1\rangle \langle 0|$. Note that $\hat A$ picks out interference terms; when
wedged between two states, $\langle \psi_1|\hat A|\psi_2\rangle$, it links the $|0\rangle$ component of one
state to the $|1\rangle $ component of the other. In fact, taking a basis of the form \begin{equation}|0\rangle =\left( \begin{array}{c}1\\
0\end{array}\right),\qquad |1\rangle =\left( \begin{array}{c}0\\ 1\end{array}\right),\end{equation} this operator is simply the first Pauli spin matrix: $\hat
A=\sigma_x =\left( \begin{array}{cc} 0 & 1\\ 1 & 0\end{array}\right)$. It is therefore easily seen that $\hat A^2 =I$, the identity
operator. Taking the expectation value (the mean value in the given state), we find \begin{eqnarray}\langle \hat A\rangle &=& \langle\psi | \hat A |\psi \rangle \; =\; \cos\phi , \\
\langle \hat A^2\rangle &=& \langle\psi | \hat A^2 |\psi \rangle \; =\; 1 .\end{eqnarray} Therefore, the uncertainty in the measurement of $\hat A$ is
\begin{equation}\Delta A = \sqrt{\langle \hat A^2\rangle
-\langle \hat A\rangle^2 } = \sqrt{\left( 1-\cos^2\phi\right)} =|\sin\phi |.\end{equation} If we repeat the experiment $N$ times on an identical ensemble of
states, standard statistical theory tells us that the uncertainty should be reduced by a factor of $\sqrt{N}$, so that $\Delta A ={{|\sin\phi |}\over \sqrt{N}}$.
Our best estimate of $\phi$ is then the value $\bar \phi =\cos^{-1}{{\langle \hat A\rangle }\over N}$, and the uncertainty in this phase estimate is given by
\begin{equation}\Delta \phi = {{\Delta A}\over {|{{d\langle \hat A\rangle }\over {d\phi}}|}} = {{\sqrt{N}|\sin\phi |}\over {N|\sin\phi |}} ={1\over
\sqrt{N}}.\end{equation}  This is the standard quantum limit or shot noise limit. In the case of single photon states being measured $N$ times, this uncertainty can be thought of
as being due to photonic shot noise or "sorting noise", the Poisson-distributed random fluctuations of the photon number in each arm of the apparatus due to the
random choice made by the beam splitter as to which way to send each photon. If, instead of $N$ single-photon states, the experimenter uses a single coherent
state pulse with mean photon number $\langle \hat N\rangle =N$, the result for the uncertainty is the same, since the random fluctuations of photon number in the
coherent state beam have the same Poisson statistics as the sorting noise.

The derivation of the standard quantum limit above assumed that the $N$ photons used to obtain the estimate were all acting independently. However, it is
possible to send in states where the photons are entangled (see appendix), so that the entire $N$-photon set is described by a single joint quantum state.  At
any horizontal position in Figure \ref{mzfig}, let a quantum state with $n_1$ photons in the top branch and $n_2$ photons in the lower branch be denoted by
$|n_1\rangle |n_2\rangle$, or more simply by $|n_1,n_2\rangle$. Then imagine sending an entangled $N$-photon state into the system; for example, a so-called NOON
state, ${1\over \sqrt{2}}\left( |N,0\rangle +|0,N\rangle \right)$ \cite{boto}. A two-photon NOON state can be easily produced by means of the Hong-Ou-Mandel
effect \cite{hom}.  $N=3$ \cite{mitchell}, $N=4$ \cite{walther}, and $N=5$ \cite{afek,israel1} states have been produced using parametric down conversion
followed by post-selection of states with the desired form. After the first beam splitter and the phase shift, the state reaching the second beam splitter is
$|\psi_N\rangle ={1\over \sqrt{2}}\left( |0,N\rangle +e^{iN\phi }|N,0\rangle \right)$ . Note that because of the entanglement, the phase shifts from the $N$
photons act collectively, giving a total phase shift $N\phi$ to one component of the entangled state. This is in contrast to the single-photon and separable
$N$-photon cases above, where each photon privately carried its own phase shift, independently of what happened to the other photons. In place of the operator
$\hat A$ above, an appropriate measurement operator to extract this total phase shift is now $\hat B_N =|0,N\rangle \langle N,0| + |N,0\rangle \langle 0,N|$. It
is straightforward to check that the expectation values and uncertainties are given by
\begin{eqnarray}\langle \hat B_N\rangle &=& \langle\psi_N | \hat B_N |\psi_N
\rangle \; =\; \cos N\phi , \label{cosnosc}\\
\langle \hat B^2_N\rangle &=& \langle\psi_N | \hat B^2_N |\psi_N \rangle \; =\; 1 \\
\Delta B_N &=& \sin N\phi\\
\Delta \phi &=&{1\over N}. \label{delphi1}\end{eqnarray} In other words, this entangled system can beat the standard quantum limit. It in fact saturates the
Heisenberg limit, the fundamental physical bound imposed by quantum mechanics \cite{lane}. The use of NOON states to carry out phase microscopy near the
Heisenberg limit has been experimentally demonstrated in \cite{israel2}.

We see that entanglement gives a ${1\over \sqrt{N}}$ advantage in phase sensitivity over unentangled $N$-photon states. The increase in oscillation frequency,
signalled by the factor of $N$ inside the cosine in Equation \ref{cosnosc} indicates super-resolution as well. Although this improved phase measurement ability
has been demonstrated in the lab, NOON states and other types of entangled photon states are difficult to create for large $N$, so one prominent current goal is
to find ways to make such states more easily, in order to allow their use in application settings outside the lab.

The quality of the phase estimation is normally quantified by the uncertainty, as obtained from some observable, as in Equation \ref{delphi1}. In more complicated
situations, such as imaging, the optimal uncertainty may be obtained via information theoretic means through the Cramer-Rao bound \cite{cramer,rao}. Alternatively, the
uncertainty may not be used at all; instead, the quality of the estimation may be measured by means of mutual information \cite{simonbahder,bahder}. Bayesian
analysis \cite{hradil,pezze} is sometimes used to provide strategies for optimizing the estimation strategy.

Other entangled $N$-photon states, such as those of \cite{yurke,huver} may be used. Some of these states give slightly better sensitivity or are slightly more
robust to noise, but overall they give qualitatively similar results to those derived from NOON states. For completeness, we should mention that entangled states
are not necessary to surpass the standard quantum limit. For example, squeezed states have been shown to be capable of achieving ${1\over {N^{3/4}}}$ phase
sensitivity \cite{caves}. However, up to this point only entangled states seem to be capable of fully reaching the Heisenberg bound. Also, it has been shown that
through a post-selection scheme \cite{resche} it is possible to produce super-resolution without the use of entangled illumination states (although
super-sensitivity was not demonstrated).

\section{Angular and rotational measurements}

\subsection{Optical orbital angular momentum} \label{oamsection}

In the last section, it was shown that phase measurements could be made with entangled optical states that exceed the standard quantum limit. It is natural to
ask if a similar improvement can be made in measurements of other quantities. In this section we show that this can be done for \emph{angular displacements}
\cite{jha}. This is done by using states for which angular rotations lead to proportional phase shifts; the super-sensitivity to phase then leads to
super-sensitivity to orientation angles. In this subsection we introduce the idea of optical angular momentum, then look at its applications in the following
subsections.

It is well known that photons carry one unit of spin angular momentum. This spin manifests itself as circular polarization: left- and right-circularly polarized
photons have spin quantum numbers $s_z=\pm 1$ along the propagation axis, while linearly polarized photons are formed from equal superpositions of the two spin
states. It is only in the past two decades that it has been widely recognized that in addition to this intrinsic spin angular momentum a photon can also carry
\emph{orbital angular momentum} (OAM) $\bm{\hat L}$ about its propagation axis \cite{allen}.  A number of excellent reviews of the subject exist, including
\cite{yao,twisted,franke}. Here, we briefly summarize the basic facts.

For a photon to have nonzero OAM, its state must have nontrivial spatial structure, such that the wavefronts have azimuthally-dependent phases of the form
$e^{i\phi(\theta )}=e^{il\theta}$. Here, $\theta$ is the angle about the $z$-axis and $l$ is a constant known as the topological charge. There are several ways
to imprint such phases onto a beam. The most common are by means of spiral phase plates (plates whose optical thickness varies azimuthally according to $\Delta
z= {{l\lambda\theta }\over {2\pi (n-1)}}$ for light of wavelength $\lambda$) \cite{beij}, computer generated holograms of forked diffraction gratings
\cite{bazh}, or spatial light modulators (SLM).

\begin{figure}
\centering
\includegraphics[scale=.3]{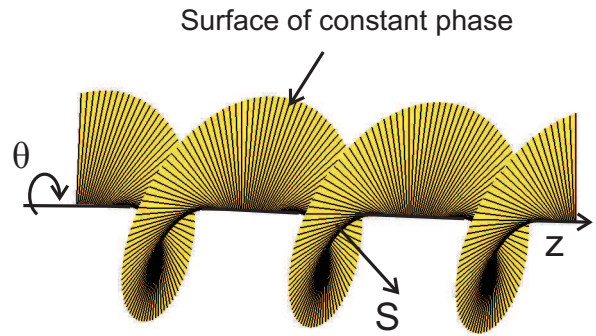}
\caption{Optical wavefronts with nonzero orbital angular momentum have corkscrew-shaped wavefronts. The Poynting vector, ${\bm S}$ must be
everywhere perpendicular to the wavefront, so it rotates as the wave propagates along the $z$-axis.}
\label{wavefrontfig}       
\end{figure}

Since the angular momentum operator about the $z$ axis is $\hat L_z=-i\hbar {\partial\over {\partial\theta}}$, the resulting wave has a $z$-component of angular
momentum given by $L_z=l\hbar$, where $\hbar$ is Planck's constant. The fact that the wavefunction must be single valued under rotations $\theta\to \theta +2\pi$ forces $l$ to be quantized to
integer values. The $e^{il\theta}$ phase factor has the effect of tilting the wavefronts, giving them a corkscrew shape (Figure \ref{wavefrontfig}). The Poynting vector $\bm S=\bm E\times \bm H$ must be
perpendicular to the wavefront, so it is at a nonzero angle to the propagation direction. $\bm S$ therefore rotates about the axis as the wave propagates,
leading to the existence of nonzero orbital angular momentum.

The question of separating the angular momentum into spin and orbital parts in a gauge-independent manner is complicated for the general case
\cite{barnettallen,vanenk1,vanenk2,zhao,nieminen,santamoto}, but as long as we restrict to the  transverse, propagating radiation field in the paraxial region, the splitting is unambiguous and OAM will be conserved in the the parametric down conversion process discussed below.

Several different optical beam modes can carry OAM; here, we focus on Laguerre-Gauss (LG) modes. (Other possibilities include higher-order Bessel or
Hermite-Gauss modes.) The LG wavefunction, or spatial amplitude function is \cite{allen2} \begin{eqnarray}& &u_{lp}(r,z,\theta )= {{C_{\rm
p}^{|l|}}\over {w(z)}}\left( {{\sqrt{2}r}\over {w(z)}}\right)^{|l|} e^{-r^2/w^2(r)}L_{\rm p}^{|l|} \left( {{2r^2}\over {w^2(r)}}\right)\nonumber \\ & &\quad
\times \; e^{ -ikr^2z/\left(2(z^2+z_R^2)\right)}e^{-i\theta l+i(2p+|l|+1)\arctan (z/z_R) },\label{lag}\end{eqnarray} with normalization $ C_{\rm p}^{|l|} = \sqrt{
{{2p!}\over {\pi (p+|l|)!}}} $ and beam radius $w(z)= w_0\sqrt{1+{z\over {z_R}}} $ at $z$. $L_{\rm p}^{\alpha}(x)$ are the associated Laguerre polynomials
\cite{arfken}. $z_r={{\pi w_0^2}\over \lambda}$ is the Rayleigh range, $w_0$ is the radius of the beam waist, and the arctangent term is the Gouy phase familiar from Gaussian laser beams. The index
$p$ characterizes the radial part of the mode; the number of intensity nodes in the radial direction is $p+1$. All LG modes (for $l\ne 0$) have a node at the central axis, about which the optical vortex circulates. For $p>1$, additional nodes appear as dark rings, concentric about the
central axis. It is important to note that not only macroscopic light beams, but even single photons can have nontrivial spatial structure leading to nonzero OAM.

Measurement of the OAM of a beam can be accomplished in a number of ways, including an interferometric arrangement that sorts different $l$ values
into different outgoing spatial modes \cite{leach1,leach2,gao} and sorting by means of q-plates \cite{karimi} (devices which couple the OAM to the polarization), as well as by  use of polarizing Sagnac interferometers \cite{slussarenko}, pinhole
arrangements followed by Fourier-transforming lenses \cite{guo}, and specialized refractive elements \cite{padgett1,berkhout1,lavery1}

As was mentioned earlier, the most common way to produce entangled photons is via spontaneous parametric down conversion (SPDC) in a $\chi^2$ nonlinear crystal. In this process, interactions with the crystal lattice cause a high frequency photon from the incident beam (called the pump beam) to split into two lower frequency photons (called, for historical reasons, the signal and idler). The signal-idler pair is entangled in a number of different variables, including energy, momentum, and polarization. Our interest here, though, is in the fact that they are entangled in OAM \cite{mair}. We will assume for simplicity that the pump beam has no OAM, in which case the signal and idler have equal and opposite values of topological charge, $\pm l$.  The $p$ values of the signal and idler are unconstrained, but since the detectors are usually coupled via fibers that propagate only $p=0$ modes, we will restrict attention to the case where the outgoing $p$ values vanish. The output of the crystal may then be expanded as a superposition of signal-idler
states with different OAM values. For the case we are considering, this takes the form
\begin{equation}\sum_{l_s,l_i =-\infty }^\infty K_{l_s
,l_i } \; \delta(l_s +l_i )\; |l_s ,l_i\rangle     = \sum_{l_s=-\infty }^\infty K_{l_s,-l_s} \;  |l_s,-l_s\rangle    ,\label{statel}\end{equation} where $l_s,l_i$ are the OAM of the signal and idler. Explicit expressions for the
expansion coefficients $K_{l_s,l_i }$ may be found in \cite{torres2,ren,barbossa,miatto}.


\subsection{Angular displacement measurement}\label{angledispsection}

\begin{figure}
\centering
\includegraphics[scale=.3]{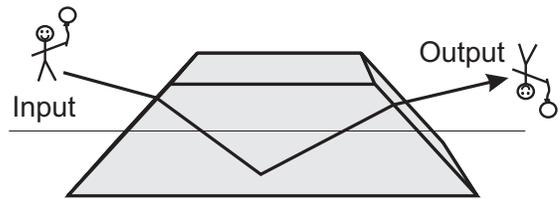}
\caption{Dove prism. Total internal reflection at the horizontal surface causes images to be inverted vertically.}
\label{dovefig}       
\end{figure}

In order to illustrate super-resolution in angular measurement, consider a Dove prism. This is a prism (Figure \ref{dovefig}) designed so that total internal
reflection occurs on the bottom surface. Because of this reflection, images are inverted in the vertical direction (the direction perpendicular to the largest
prism face), without any corresponding horizontal inversion. One interesting property of these prisms is that if they are rotated by an angle $\theta$ about the
propagation axis, the image is rotated by $2\theta$. For well-collimated light with a definite value of orbital angular momentum, $L=l\hbar$, the Dove prism
reverses the angular momentum direction, while maintaining its absolute value: $l\to -l$, assuming that the beam is not too tightly focused \cite{gonzalez}.

\begin{figure}
\centering
\includegraphics[scale=.35]{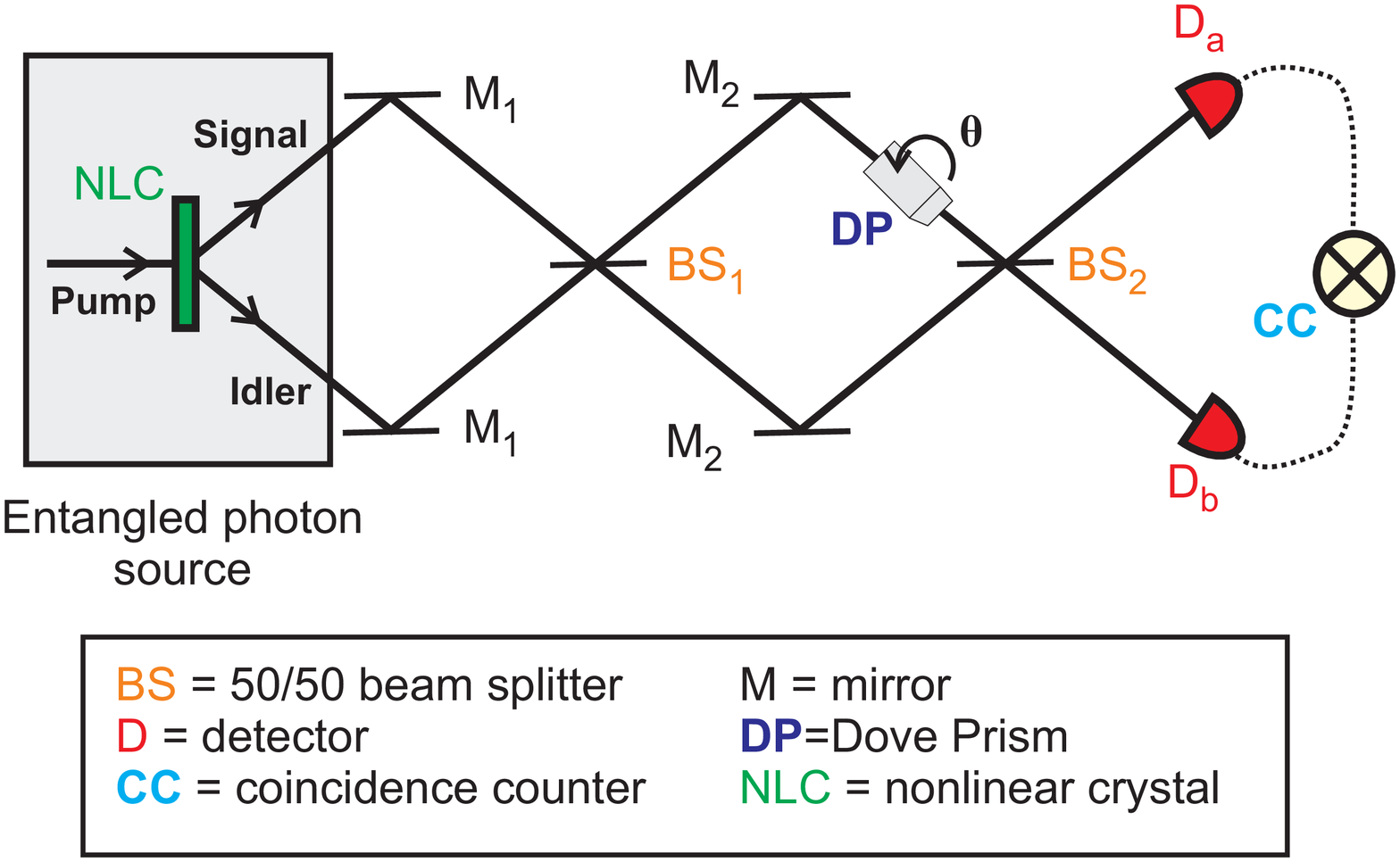}
\caption{Schematic of setup for super-resolution angular displacement measurements using entangled OAM states.
Spontaneous parametric down conversion in a nonlinear crystal (at left) sends entangled photon pairs into a Mach-Zehnder interferometer (middle portion)
which contains a Dove prism in one arm. At the right, photodetectors are connected to a coincidence circuit, so that an event is registered only when both
detectors fire within a short time window.}
\label{jhafig}       
\end{figure}

Consider the setup shown in Figure \ref{jhafig} \cite{jha}. The Dove prism is rotated by some unknown angle $\theta$ and the goal is to measure
$\theta$ as precisely as possible. As input, take a pair of entangled photons, with signal and idler of topological quantum numbers $+l$ and $-l$,
respectively.
The state produced by the entangled light source will be of the form of Equation \ref{statel}, but insertion of appropriate filters can be used to block all except
the $\pm l$ terms for some fixed value of $l$:
\begin{equation}|\psi_{in}\rangle={1\over\sqrt{2}} \left( |l\rangle_s |-l\rangle_i +|-l\rangle_s |l\rangle_i\right) ,\end{equation} where $s$ and $i$ denote signal and
idler, which we take to enter the interferometer in the upper and lower branches respectively.
The effect of the components of the apparatus on the two-photon state can be traced through to find the output state, $|\psi_{out}\rangle$.
The rate of detection events for which value $l$ arrives one detector and value $-l$ arrives the other  is then the expectation value of the operator \begin{eqnarray}\hat R&=&
|+l\rangle_{D_b}|-l\rangle_{D_a} \cdot_{D_a} \langle+l|_{D_b}\langle-l| \\ & & \qquad +|-l\rangle_{D_b}|+l\rangle_{D_a} \cdot_{D_a} \langle -l|_{D_b}\langle +l|.\nonumber\end{eqnarray} We find the expectation values
\begin{eqnarray}\langle \hat R\rangle &=& \langle \psi_{out}|\hat R|\psi_{out}\rangle =\cos^2(2l\theta), \\
\langle \hat R^2\rangle&=& \langle \psi_{out}|\hat R^2 |\psi_{out}\rangle =\cos^2(2l\theta) ,\end{eqnarray} so that the uncertainty in $\hat R$ is
\begin{equation} \Delta \hat R = \sqrt{\langle \hat R^2\rangle -\langle \hat R\rangle^2} =\cos (2l\theta) \sin (2l\theta) ={1\over 2}\sin(  4l\theta) .\end{equation}
We may then extract an estimate of the angular displacement from the value of $\langle \hat R\rangle $. The resulting estimate will have uncertainty
\begin{equation}\Delta \theta = {{\Delta \hat R}\over {|{{\partial \langle \hat R\rangle }\over {\partial \theta}}|} }. \end{equation}

This whole procedure can be generalized from two to $N$ entangled photons. In an identical manner it can be shown that the phase sensitivity is now given by
\begin{equation}\Delta \theta = {1\over {2Nl}}. \end{equation} This should compared to the case when an ensemble of $N$ independent photons are measured with the
same apparatus: \begin{equation}\Delta \theta_{SQL} = {1\over {2\sqrt{N}l}}. \end{equation} We not only see once again that the entanglement allows the
sensitivity to be reduced by a factor of $\sqrt{N}$ beyond the case standard quantum limit, but further improvements can be made to the angular sensitivity by
using states of higher angular momentum. The angular resolution can also exceed the classical limit, with the argument of the oscillatory term in $\langle \hat
R\rangle$ being proportional to $Nl\theta$. Such high precision measurement of rotations may ultimately be useful in nanotechnology applications.

\subsection{Image reconstruction and object identification}

An additional set of OAM-based approaches illustrates the idea that single photon states can be chosen specifically to suit the type of measurement that will be
made with them. The Laguerre-Gauss modes $u_{lp}(r,z,\theta )$ form a complete set, meaning that any two-dimensional function can be built as a linear
combination of them: $f(r,\theta ) =\sum_{l,p} a_{lp}u_{lp}(r,z,\theta )$, for some set of coefficients $a_{lp}$. This fact leads to the idea of \emph{digital spiral
imaging} \cite{torner,molina,torres}. This is a form of angular momentum spectroscopy in which a beam of known OAM is used to illuminate an object, and then the
spectrum of outgoing OAM values  is measured after reflection or transmission from the object. This allows the evaluation of the squared coefficients
$|a_{lp}|^2$ of the object's reflection or transmission profile, and reveals a great deal of information about the object structure.  In this manner, the object
could be identified from a known set, or its difference from some desired shape could be quantified. This approach has two advantages: first, for simple objects
where only a small number of coefficients are large, the OAM spectrum can be determined to good accuracy with a small number of photons. Second, rotating the
object only adds a constant phase factor to all the amplitudes, leaving the intensities of the outgoing components unchanged. This means that the spatial
orientation of the object does not need to be known, and that the object can still be easily identified even if it is rapidly rotating \cite{fitz}. Measurement
of the OAM spectrum also allows for rapid experimental determination of the rotational symmetries of an unknown object \cite{simonspiral,uribe}. These facts
could make this approach ideal for a number of applications, such as (i) looking for sickle cells, cancer cells, or other irregularly-shaped cells in a moving
bloodstream, (ii) rapidly scanning for defects in parts on an assembly line, or (iii) identifying defects in rapidly spinning objects, such as a rotor blade.

One goal of this approach is to ultimately be able to reconstruct the image from the OAM spectrum alone. Two problems must be addressed in order to do this. The
first is that measuring the intensities of the outgoing OAM intensity components is not sufficient for image reconstruction: the coefficients $a_{lm}$ are
complex, so that their relative phases must be determined as well. This, fortunately, can once again be done by means of an interferometric arrangement such as
that of Figure \ref{mzfig} \cite{fitz,simonspiral,uribe}; this method has been called \emph{correlated spiral imaging}. The second problem is that current technology
does not easily allow terms with nonzero $p$ values to be measured efficiently at the single photon level. So even though the angular structure of the object can
be reconstructed, nontrivial radial structure is difficult to detect in this manner. This will likely be remedied over time as improved detection methods are
developed.

Digital and correlated spiral imaging do not require entanglement for their basic functioning. However, it is clear that $N$-photon entangled states would once
again allow these approaches to exceed the classical limits on resolution and sensitivity in the same manner as the other methods described in the earlier
sections. In particular, angular oscillation frequencies would again be increased by a factor of $N$ as in the previous subsection, allowing the same improvement
in sensitivity and resolution of small angular structures. In this way, super-resolved imaging could in principle be carried out entirely by OAM measurements.

\subsection{Rotational measurements}

Since measurements with OAM are being discussed, one more related type of measurement can be mentioned, even though it does not require entanglement or low photon numbers.
Optical orbital angular momentum states can also be used to measure rotation rates \cite{vasnetsov,lavery3,rosales,lavery4,padgettphystod}. Imagine two photons of opposite OAM $\pm l$ reflecting off the surface of a rotating object. Let $\omega$ be the photon frequency and $\Omega $ be the rotational frequency of the object. The two reflected photons will experience equal and opposite Doppler shifts: $\omega\to \omega \pm l\Omega$. If the two reflected photons are allowed to interfere, the combined intensity will therefore exhibit a beat frequency of $2l\Omega$. If $l$ is large, then precise measurements can be made of even very small values of $\Omega$.

\section{Frequency, Polarization, and Dispersion}

Here, we briefly mention several other applications, taking a similar approach to those discussed in the earlier sections.

\subsection{Frequency Measurement}

\begin{figure}
\centering
\includegraphics[scale=.25]{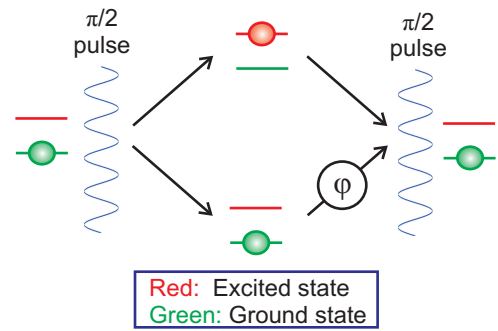}
\caption{Schematic of a Ramsey interferometer. The first pulse puts the ground state input into a superpositions of ground and excited states. Free propagation causes the two terms in the superposition to differ in phase by an amount $\phi$ proportional to the transition frequency. }
\label{ramfig}       
\end{figure}

The Ramsey interferometer, used to measure atomic transition frequencies, is formally equivalent to a Mach-Zehnder interferometer. A schematic of the Ramsey
interferometer is shown in Figure \ref{ramfig}. A $\pi\over 2$ pulse is used to put an incoming atom into a superposition of the ground and excited states; this
is analogous to the first beam splitter in the Mach-Zehnder case, which puts a photon into a superposition of two different paths. The two energy states gain
different phase shifts under free propagation for a fixed time, so that when they are returned to the same state again, they will produce an interference pattern
as the time between the two pulses is varied. The relative phase shift is ${{\Delta E\; t}\over \hbar}=\omega t$, where $\Delta E$ and $\omega$ are the energy
difference and emitted photon frequency of the levels, $\hbar$ is Planck's constant, and $t$ is the time between the two pulses. As a result, the interference
pattern allows determination of the frequencies in a manner exactly analogous to the phase determination in a Mach-Zehnder interferometer. The formal analogy
between the Mach-Zehnder and Ramsey interferometers (as well as with quantum computer circuits) has been referred to as the quantum Rosetta stone
\cite{kapale,lee2}, as it allows ideas from one area to be translated directly into the other.

Because of the exact mathematical equivalence between the two interferometers, it follows immediately that the use of entangled atomic states as input to the
Ramsey interferometer should allow super-sensitive measurements of frequency, with the uncertainty scaling as  ${1\over N}$ as the number $N$ of entangled atoms
increases \cite{bollinger,bouyer}. In a similar manner, Heisenberg-limited resolution can be achieved in optical and atomic gyroscopes, which are also based on
Mach-Zehnder interferometry \cite{dowlinggyro}.

\subsection{Dispersion}

Dispersion, the differential propagation speed of different states of light, must be taken into account to very high precision in modern optical
telecommunication systems, in long-range stand-off sensing, and in many other areas. Here, we briefly mention several related quantum approaches to the
measurement and cancelation of dispersion.

Chromatic dispersion effects can be described by considering the wavenumber $k$ to be a function of frequency: $k(\omega )$. In a dispersive medium, $k$ will be
a nonlinear function of $\omega$, causing different frequencies to propagate at different speeds. This leads to broadening of wave packets as they propagate,
reducing the resolution of measurements made with those packets. Expanding $k(\omega )$ in a Taylor series, it can be divided into terms that are even-order and
terms of odd-order in frequency. It was found in the early 1990's that the use of an entangled light source combined with coincidence counting can cause the even
order terms to cancel; among other things, this means that the largest contribution to the broadening (the second-order group delay term), will be absent. Unlike
the applications in the previous section, this method does not require simultaneous entanglement of a large number $N$ of photons: it only requires two photons
to be entangled at a time, and so can be easily carried out with standard down conversion sources. These sources naturally produce photon pairs with frequencies
$\omega_0 \pm \Delta \omega$ displaced by equal amounts about the central frequency, $\omega_0$. ($\omega_0$ is half the pump frequency.) The two photons are in
fact entangled in the frequency variable: it is not predetermined which photon has the higher frequency and which has the lower, so both possibilities have to be
superposed, \begin{eqnarray}|\psi\rangle &=& {1\over \sqrt{2}}\left( |\omega_0+\Delta \omega\rangle_{signal} |\omega_0-\Delta \omega\rangle_{idler}\right. \\ & &
\qquad  + \left. |\omega_0-\Delta \omega\rangle_{signal} |\omega_0+\Delta \omega\rangle_{idler}\right) .\nonumber\end{eqnarray}

For even-order dispersion cancelation, the essential idea is make the photons interfere in such a way that the two opposite-signed frequency displacements add,
causing the even powers in the expansion to cancel. There are two ways to do this. The Steinberg-Kwiat-Chaio (SKC) method \cite{steinberg1,steinberg2} puts the
dispersive material into one arm of a Mach-Zehnder interferometer, sends the entangled photons into the system at one end, and then measures coincidence counts
in the two output ports at the other end. The Franson method \cite{franson}, on the other hand sends the two photons into two materials with opposite-dispersion,
looking again at coincidence counts at the output. Either way, the second order pulse-broadening terms are absent.

It has since been shown that the SKC method can be reproduced with classical light sources (no entanglement) \cite{erkmen1,resch2}, but the Franson method seems
to truly require entanglement. It has also been shown \cite{minaeva} that by concatenating two successive interferometers (one Hong-Ou-Mandel interferometer
\cite{hom} and one Mach-Zehnder), more complicated manipulations of the even- and odd-order frequency dispersion terms can be carried out, such as canceling the
odd-order terms or causing even-order and odd-order cancelation to occur simultaneously in different parts of an interferogram.

Optical coherence tomography (OCT) is a well-known interferometric method for detecting and mapping subsurface structures in biological samples. It makes use of first-order amplitude-amplitude correlations. \emph{Quantum optical coherence tomography }(QOCT) \cite{abouraddy,nasr1,nasr2} is a similar method that makes use of entangled photon sources and second-order intensity-intensity correlations. QOCT has greatly improved resolution due to the dispersion cancelation described above. The main difficulty with the method is that it is very slow due to the need for individual pairs of entangled photons. Several variations \cite{erkmen1,resch2,bana,kalten,legouet} have been proposed that mimic the dispersion-canceling effect of QOCT to varying degrees using classical light sources, including an experimental demonstration of dispersion-cancelled OCT with a $15$-fold reduction in the broadening of the interference peaks \cite{mazurek}.

In addition to chromatic dispersion, optical pulses can also exhibit polarization mode dispersion (PMD) in which components of an optical pulse with different polarizations travel at different speeds. With the extensive use of
reconfigurable optical add-drop multiplexer (ROADM) systems in modern fiber optics networks it has become essential to be able to characterize the PMD of small
components such as wavelength selective switches to very high levels of precision. Quantum interferometric methods similar to those used for chromatic dispersion
cancelation have also been shown to allow ultra-high precision PMD measurements \cite{branning,dauler,fraine1,fraine2}, beyond those possible using
interferometry with classical states of light. In particular, the apparatus of \cite{fraine1} has been demonstrated to produce time delay measurements with
sensitivities as low as the attosecond ($10^{-18}$ s) level \cite{fraine2} as a result of even-order dispersion cancelation; this was the first demonstration of a quantum metrology application implemented on an
industrial commercial device.
The apparatus for super-resolved PMD measurement is once again conceptually similar to the interferometers of Figures \ref{mzfig}, expect that now the first beam splitter sends different polarizations into different spatial modes, the phase shift in one arm is introduced by dispersion, and the two polarizations are mixed at the end by diagonally-oriented polarizers instead of by a second beam splitter.

\subsection{Quantum Lithography}

Although not a sensing or measuring application, one other topic should be mentioned which is closely related to the subjects of this paper. That is the idea of
quantum lithography \cite{boto,dangelo,bjork1,bjork2}, in which entangled states such as NOON states are used to write subwavelength structures onto a substrate.
This again is due to the fact that the $N$-photon entangled state oscillates $N$ times faster than the single-photon state, or equivalently, that the effective
wavelength of the system is reduced from $\lambda$ to $\lambda\over \sqrt{N}$. The method holds great promise, in principle, for use in fabricating more compact
microchips and for other nanotechnology applications. However, practical application of the idea is again currently hampered by the difficulty of creating high
$N$ entangled states on demand. Moreover, recent work \cite{kothe} has shown that the efficiency of the process is low for small values of $N$.

\section{Conclusions}

The methods of quantum sensing described in this review (use of entangled states, choosing single-photon or multi-photon states that are tailored to optimize the desired measurement, and use of intensity-intensity correlations), in principle allow great advantages over classical methods. The current challenges involve overcoming the limitations that arise in real-life practice. Chief among these challenges are:

$\bullet$ Producing high-quality entangled states with $N>2$ photons is very difficult to arrange, and the efficiencies of current methods are extremely low.

$\bullet$ For more sophisticated OAM-based applications, optical fibers that propagate multiple $l$ values and an effective means of measuring the radial $p$ quantum number are needed.

$\bullet$ NOON states are fragile against loss and noise \cite{huver}, so they are not suitable for use in noisy real-world environments. However, methods have been developed to find other states with more robust entanglement \cite{huver,dorner,demkow,lee}. Some of these states have been shown to beat the standard quantum limit, and there is hope for ultimately finding states that strongly approach the Heisenberg limit in lossy and noisy environments.

If these technical challenges can be met, quantum measurement methods will be ready to move from the laboratory to the real world, opening a wide range of new
applications and improved measurement capabilities.

%
%

\section*{Appendix: Brief Review of Quantum Mechanics}

As the size scales involved in a measurement or as the number of particles involved becomes small, the laws of classical physics break down and the formalism of
quantum mechanics must be taken into account. There are many detailed introductions to quantum mechanics from various points of view (for example \cite{shankar,peres,griffiths}). Here, we
give just the few minimal facts needed for this review.

The state of a particle (or more generally, of any quantum system) is described by a vector in a complex space, known as a Hilbert space. In Dirac notation,
these vectors are written as $|\psi\rangle$, $|\phi\rangle$, etc., where $\psi$, $\phi$, are just labels for the states. One particularly useful example is the Fock state $|n\rangle$, which contains exactly $n$ photons.  The operators $\hat a^\dagger$ and $\hat a$, known as creation and annihilation operators, raise and lower the number of photons in the Fock state: $\hat a^\dagger |n\rangle = \sqrt{n+1}| n+1\rangle$, and $\hat a |n\rangle = \sqrt{n}| n-1\rangle$. The number operator, $\hat N=\hat a^\dagger \hat a$, counts the number of photons in the state: $\hat N|n\rangle =n|n\rangle$.

To each vector $|\psi \rangle $ there is associated a corresponding dual or conjugate vector $\langle \psi |$. This dual vector is the Hermitian conjugate (the complex transpose) of the original vector. So, for example, if $|\psi\rangle$ is described by a column vector with components $\psi_1,\psi_2,\dots$, then $\langle \psi |$
is a row vector with components $\psi_1^\ast,\psi_2^\ast ,\dots$:
\begin{equation}|\psi\rangle =  \left( \begin{array}{c} \psi_1\\ \psi_2\\ \dots \end{array} \right)  \quad \leftrightarrow \quad  \langle \psi | =
\left( \psi_1^\ast \;  \psi_2^\ast \; \dots \right)       . \end{equation} Inner products are defined between bras and kets: $\langle \phi |\psi\rangle
=\phi_1^\ast \psi_1 + \phi_2^\ast \psi_2+\phi_3^\ast \psi_3+\dots $. All vectors are usually assumed to be normalized: $\langle \psi |\psi\rangle =1$. Inner
products such as $|\phi\rangle \langle \psi|$ act as operators that project state $|\psi\rangle$ onto the direction parallel to state $|\phi\rangle$.

One of the defining properties of quantum mechanics is that it is a linear theory: the dynamics are defined by the Schr\"odinger equation, a linear
second-order partial differential equation. As a result of this linearity, quantum systems obey the superposition principle: any two possible states of a system,
say $|\psi\rangle$ and $ |\phi\rangle$ can be added to get another allowed state of the system: $|\Phi\rangle ={1\over {\sqrt{2}}}\left( |\psi\rangle +
|\phi\rangle\right)$, where the ${1\over {\sqrt{2}}}$ is included to maintain normalization. If $|\psi\rangle$ and $|\phi\rangle$ are not orthogonal to each
other, $\langle \phi |\psi\rangle \ne 0$, then the two terms in the superposition may interfere with each other, leading to many uniquely quantum
phenomena.

Suppose a pair of particles, $A$ and $B$, together form a composite system, $C$. If $A$ is in state $|\psi\rangle$ and $B$ is in state $|\phi\rangle$, then the
composite system is in state $|\Psi\rangle_C =|\psi\rangle_A |\phi\rangle_B$, where the subscripts are used to indicate which system is in which state. Such a
state is called a product state or a \emph{separable} state. Often, however, the state of the composite system is known while the states of the individual
subsystems are not; in that case, all the possibilities consistent with the available information have to be superposed, as for example in the state:
$|\Phi\rangle_C ={1\over {\sqrt{2}}} \left( |\psi\rangle_A |\phi\rangle_B + |\phi\rangle_A |\psi\rangle_B \right)$. The two possibilities ($A$ in state
$|\psi\rangle$ with $B$ in state $|\phi\rangle $, versus $A$ in state $|\phi\rangle$ with $B$ in state $|\psi\rangle $), can be thought of as {\it both} existing
simultaneously. Such a state, which cannot be factored into a single well-defined state for $A$ and a state for $B$ is called \emph{entangled}.  An example is
the case where two photons arrive at a beam splitter: both can exit out one output port, both can exit at the other output port, or there are two ways in which the two photons can exit at different ports. If the locations of the outgoing photons are not measured, then there is no way to distinguish between the possibilities and they
must all be included: the state of the full system is therefore an entangled state formed by the superposition of all four possibilities. If a measurement is
made that determines the states of one of the subsystems, then the entangled state collapses to a product state, and the state of the second subsystem is also therefore known. For example if particle $A$ is measured and
found to be in state $\psi$, then the collapse $|\Phi\rangle_C ={1\over {\sqrt{2}}} \left( |\psi\rangle_A |\phi\rangle_B +|\phi\rangle_A |\psi\rangle_B \right)
\to |\psi\rangle_A |\phi\rangle_B $ occurs. After the collapse, we know that $B$ must now be in state $\phi$. It is important to point out that the subsystems were not
in definite states $|\psi\rangle_A$ and $ |\phi\rangle_B$ before the measurement, but that their prior states as individuals did not even exist. It can be shown that
the contrary assumption of prior existence of definite states for each subsystem before measurement leads to contradictions with experiment, as demonstrated by the
violation of Bell and CHSH inequalities \cite{bell1,bell2,clauser1,clauser2,aspect}. The fact that entangled subsystems do not exist in definite states before
measurement is one of the odder and more non-intuitive aspects of quantum mechanics, but is extremely well verified experimentally.

The most common way to produce entangled pairs of photons is via spontaneous parametric down conversion (SPDC) \cite{shih}, in which nonlinear interactions in a crystal mediate the conversion of an incoming photon (the \emph{pump}) into two lower energy outgoing photons (the \emph{signal and idler}). Only a tiny fraction of the incoming pump photons undergo down conversion (typically on the order of one in $10^6$), but SPDC is very versatile in the sense that the down converted photons can be entangled in multiple different variables (frequency, time, angular momentum, linear momentum, and polarization), and the properties of the entangled pairs can be finely tuned by proper choice of parameters for the nonlinear crystal and the laser beam pump.
Other sources, such as quantum dots, atomic cascades, and specially designed nonlinear optical fibers also exist; see \cite{edamatsu2} for a review.

Operators are mathematical objects that perform actions on states; these operators can include, for example, matrices and derivatives.
Quantities that can be physically measured, such as energy and angular momentum, are eigenvalues of Hermitian operators (operators which
equal their Hermitian conjugate). Two operators $\hat A$ and $\hat B$ which do not commute, $[\hat A ,\hat B]\equiv \hat A \hat B - \hat B \hat A\ne 0$, obey an
uncertainty relation: if $a$ and $b$ are the variables associated to the two operators, then there is a minimum value to the product of their uncertainties:
$\Delta a \Delta b$ must exceed a minimum quantity proportional to the commutator $[\hat A ,\hat B]$. The most famous example is
between position ($x$) and momentum ($p$). $x$ and $p$ form a conjugate pair, obeying the so-called canonical commutation relations:  $\left[ \hat x,\hat p\right]=i\hbar,$ where $\hbar $ is Planck's constant. As a result, we find the Heisenberg uncertainty relation \begin{equation}\Delta x\Delta p\ge {\hbar\over 2} .\end{equation}  Energy and time obey a similar relation, \begin{equation}\Delta E\Delta
t\ge {\hbar\over 2} .\end{equation}

These uncertainty relations provide the ultimate fundamental physical limit to all
measurements, and are the basis of the Heisenberg limit given in the main text.

\section*{Conflict of Interests}

The author declares that there is no conflict of interests regarding the publication of this paper.



\begin{thebibliography}{99}

\bibitem{lukosz} W. Lukosz, "Optical systems with resolving powers exceeding the classical limit", \emph{Journal of Optical Society America}, vol. 56, no. 11, pp. 1463-1472, 1966.

\bibitem{heintzmann} R. Heintzmann, M. G. L. Gustafsson, ``Subdiffraction resolution in continuous samples'', \emph{Nature Photonics}, vol. 3, no. 7, pp.
    362-364, 2009.

\bibitem{jacobson} J. Jacobson, G. Bjork, I. Chuang, Y. Yamamoto, ``Photonic de Broglie Waves'', {\it Physical Review Letters}, vol. 74, no. 24, pp. 4835-4838, 1995.

\bibitem{fonseca} E. J. S. Fonseca, C. H. Monken, and S. P\'adua, ``Measurement of the de Broglie Wavelength of a Multiphoton Wave Packet'', \emph{Physical Review
    Letters}, vol. 82, no. 14, pp. 2868-2871, 1999.

\bibitem{edamatsu} K. Edamatsu, R. Shimizu, and T. Itoh, ``Measurement of the Photonic de BroglieWavelength of Entangled Photon Pairs Generated by Spontaneous
    Parametric Down-Conversion'', \emph{Physical Review Letters}, vol. 89, no. 21, pp. 213601 (4 pages), 2002.

\bibitem{mitchell} M. W. Mitchell, J. S. Lundeen, A. M. Steinberg, ``Super-resolving phase measurements with a multiphoton entangled state'', \emph{Nature},
    vol. 429, no. 6988, pp. 161-164, 2004.

\bibitem{walther} P. Walther, J. W. Pan, M. Aspelmeyer, R. Ursin, S. Gasparoni, A. Zeilinger, ``De Broglie wavelength of a non-local four-photon
    state'', \emph{Nature}, vol. 429, no. 6988, pp. 158-161, 2004.

\bibitem{resche} K. J. Resch, K. L. Pregnell, R. Prevedel, A. Gilchrist, G. Journal of Pryde, J. L. O�Brien, and A. G. White, ``Time-Reversal and Super-Resolving Phase
    Measurements'', \emph{Physical Review Letters}, vol. 98, no. 22, pp.
    223601 (4 pages), 2007.

\bibitem{giov1} V. Giovannetti, S. Lloyd, L. Maccone, ``Advances in quantum metrology'', {\it Nature Photonics}, vol. 5, no. 4, pp. 222-229, 2011.

\bibitem{shih} Y. Shih, ``Entangled biphoton source�property and preparation'', \emph{Reports on Progress Physics}, vol. 66, pp. 1009-1044, 2003.

\bibitem{mandel}  L. Mandel, E. Wolf, {\it Optical Coherence and Quantum Optics}, Cambridge University Press, Cambridge, UK 1995.

\bibitem{loudon} R. Loudon, {\it The Quantum Theory of Light, 3rd ed.}, Oxford University Press, Oxford, UK, 2000.

\bibitem{clauser1} J. F. Clauser, M. A. Horne, ``Experimental consequences of objective local theories'', \emph{Physical Review D}, vol. 10, no. 2, pp. 526-535,
    1974.

\bibitem{clauser2} J. F. Clauser, A. Shimony, ``Bell's theorem: experimental tests and implications'', \emph{Reports on Progress in Physics}, vol. 41, no. 12, pp.
    1881-1927, 1978.

\bibitem{obrien} J. L. O�Brien, A. Furusawa, and J. Vu\v{c}kovi\'{c}, ``Photonic Quantum Technologies'', {\it Nature Photonics}, vol. 3, no. 12, pp. 687-695,
    2009.

\bibitem{greggsasha} A. V. Sergienko and G. S. Jaeger, ``Quantum information processing and precise optical measurement
with entangled-photon pairs'', {\it Contemporary Physics}, vol. 44, no. 4, pp. 341-356, 2003.

\bibitem{kapale} K. T. Kapale, L. D. Didomenico, H. Lee P. Kok, and J. P. Dowling, ``Quantum Interferometric Sensors'', {\it Concepts of Physics}, vol. 2, no.
    3-4, pp.
    225-240, 2005.

\bibitem{giov2} V. Giovannetti, S. Lloyd, L. Maccone, ``Advances in quantum metrology'', {\it Science}, vol. 306, no. 5700, pp. 1330-1336, 2004.

\bibitem{fraine1} A. Fraine, D. S. Simon, O. Minaeva, R. Egorov, and A. V. Sergienko, ``Precise evaluation of polarization mode dispersion by separation of even-
    and odd-order effects in quantum interferometry'', {\it Optics Express}, vol. 19, no. 23, pp. 22820-22836, 2011.

\bibitem{fraine2} A. Fraine, O. Minaeva, D. S. Simon, R. Egorov, and A. V. Sergienko, ``Evaluation of polarization mode dispersion in a telecommunication
    wavelength selective switch using quantum interferometry'', {\it Optics Express}, vol. 20, no. 3, pp. 2025-2033, 2012.

\bibitem{hradil2} Z. Hradil, ``Phase measurement in quantum optics'', {\it Quantum Optics}, vol. 4, no. 2, pp. 93-108, 1992.

\bibitem{sussglo} L. Susskind, J. Glogower, ``Quantum mechanical phase and time operator'', {\it Physics}, vol. 1, no. 1, pp. 49-61, 1964.

\bibitem{carr} P. Carruthers, M. M. Nieto, ``Phase and Angle Variables in Quantum Mechanics'', \emph{Reviews of Modern Physics}, vol. 40, no. 2, pp. 411-440, 1968.

\bibitem{levy} J.-M. Levy-Leblond, ``Who is afraid of nonhermitian operators? A quantum description of angle and phase'', \emph{Annals of Physics (N.Y.)}, vol. 101, no. 1, pp. 319-341, 1976.

\bibitem{pegg1} D. T. Pegg and S. M. Barnett, ``Phase properties of the quantized single-mode electromagnetic field'', \emph{Physical Review A}, vol. 39, no. 4, pp. 1665-1675, 1989.

\bibitem{vaccarro1}  J. A. Vaccaro and D. T. Pegg, ``Phase properties of squeezed states of light'', \emph{Optics Communications}, vol. 70, no. 6, pp. 529-534, 1989.

\bibitem{barnett1} S. M. Barnett, S. Stenholm, and D. T. Pegg, ``A new approach to optical phase diffusion'', \emph{Optics Communications}, vol. 73, no. 4, pp. 314-318, 1989.

\bibitem{vaccarro2} J. A. Vaccaro and D. T. Pegg, ``Physical Number-phase Intelligent and Minimum-uncertainty States of Light'', \emph{Journal of Modern Optics}, vol. 37, no. 1, pp. 17-39, 1990.

\bibitem{barnett2} S. M. Barnett and D. T. Pegg, , ``Quantum theory of rotation angles'', \emph{Physical Review A}, vol. 41, no. 7, pp. 3427-3435 1990.

\bibitem{vacarro3} J. A. Vaccaro and D. T. Pegg, ``Wigner function for number and phase'', \emph{Physical Review A}, vol. 41, no. 9, pp. 5156-5163, 1990.

\bibitem{summy} G. S. Summy and D. T. Pegg, ``Phase optimized quantum states of light'', \emph{Optics Communications}, vol. 77, no. 1, pp. 75-79, 1990.

\bibitem{barnett3} S. M. Barnett and D. T. Pegg, ``Quantum theory of optical phase correlations'', \emph{Physical Review A}, vol. 42, no. 11, pp. 6713-6720,
    1990.

\bibitem{pegg2} D. T. Pegg, J. A. Vaccaro, and S. M. Barnett, ``Quantum-optical Phase and Canonical Conjugation'', \emph{Journal of Modern Optics}, vol. 37, no. 11, pp. 1703-1710, 1990.

\bibitem{bergouenglert} J. Bergou, B.-G. Englert, ``Operators of the phase. Fundamentals'', \emph{Annals of Physics (N.Y.)}, vol. 209, no. 2, pp. 479-505, 1991.

\bibitem{boto} A. N. Boto, P. Kok, D. S. Abrams, S. L. Braunstein, C. P. Williams, J. P. Dowling, ``Quantum Interferometric Optical Lithography: Exploiting
    Entanglement to Beat the Diffraction Limit'', {\it Physical Review Letters}, vol. 85, no. 13, pp. 2733-2736, 2000.

\bibitem{hom} C. K. Hong, Z. Y. Ou, and L. Mandel, ``Measurement of Subpicosecond Time Intervals between Two Photons by Interference'', {\it Physical Review Letters}
    vol. 59, no. 18, pp. 2044-2046, 1987.

\bibitem{afek} I. Afek, O. Ambar, Y. Silberberg, ``High-NOON States by Mixing Quantum and Classical Light'', {\it Science}, vol. 328, no. 5980, pp. 879-881,
    2010.

\bibitem{israel1} Y. Israel, I. Afek, S. Rosen, O. Ambar, and Y. Silberberg, ``Experimental tomography of NOON states with large photon numbers'', {\it Physical
    Review A}, vol. 85, no. 2, pp. 022115 (5 pages), 2012.

\bibitem{lane} A. S. Lane, S. L. Braunstein, C. M. Caves, ``Maximum-likelihood statistics of multiple quantum phase measurements'', {\it Physical Review A}, vol. 47,
    no. 3, pp. 1667-1696, 1993.

\bibitem{israel2} Y. Israel, S. Rosen, Y. Silberberg, ``Supersensitive Polarization Microscopy Using NOON States of Light'', {\it Physical Review Letters}, vol. 112,
    no. 10, pp. 103604 (4 pages), 2014.

\bibitem{cramer} H. Cramer, \emph{Mathematical Methods of Statistics}, Princeton University Press, Princeton, NJ, USA, 1946.

\bibitem{rao} C. R. Rao, \emph{Bulletin of the Calcutta Mathematical Society}, vol. 37, no. 3, pp. 81-91, 1945.

\bibitem{simonbahder} D. S. Simon, A. V. Sergienko, T. B. Bahder, ``Dispersion and fidelity in quantum interferometry'', \emph{Physical Review A}, vol. 78,
    no. 5, pp.
    053829 (12 pages), 2008.

\bibitem{bahder} T. B. Bahder, ``Phase estimation with nonunitary interferometers: Information as a metric'', \emph{Physical Review A}, vol. 83, no. 5, pp. 053601
    (14 pages), 2011.

\bibitem{hradil} Z. Hradil, R. My\v{s}ka, J. Pe\v{r}ina, M. Zawisky, Y. Hasegawa, and H. Rauch, ``Quantum Phase in Interferometry'', \emph{Physical Review Letters},
    vol. 76, no. 23, pp. 4295-4298, 1996.

\bibitem{pezze} L. Pezz\'e, A. Smerzi, G. Khoury, J. F. Hodelin, D. Bouwmeester, {\it Physical Review Letters}, vol. 99, no. 22, pp. 223602 (4 pages), 2007.

\bibitem{yurke} B. Yurke, S. L. McCall, J. R. Klauder, ``Phase Detection at the Quantum Limit with Multiphoton Mach-Zehnder Interferometry'', {\it Physical Review A},
    vol. 33, no. 22, pp. 4033 (4 pages), 1986.

\bibitem{huver} S. D. Huver, C. F. Wildfeur, J. P. Dowling, ``Entangled Fock State for Robust Quantum Optical Metrology, Imaging, and Sensing'', {\it Physical Review
    A}, vol. 78, no. 6, pp. 063828 (5 pages), 2008.

\bibitem{caves} C. M. Caves, �Quantum mechanical noise in an interferometer,� \emph{Physical Review D}, vol. 23, no. 8, pp. 1693-1708, 1981.

\bibitem{jha} A. K. Jha, G. S. Agarwal, R. W. Boyd, ``Supersensitive measurement of angular displacements using entangled photons'', {\it Physical Review A}, vol. 83,
    no. 5, pp. 053829 (7 pages), 2011.

\bibitem{allen} L. Allen, M. W. Beijersbergen, R. J. C. Spreeuw, and J. P. Woerdman, ``Orbital angular-momentum of light and the transformation of
    Laguerre-Gaussian laser modes'', \emph{Physical Review A}, vol. 45, no. 11, pp. 8185-8189, 1992.

\bibitem{yao} A. M. Yao and M. J. Padgett, ``Orbital angular momentum: origins, behavior and applications'', \emph{Advances in Optics and Photonics}, vol. 3, no. 2, pp. 161-204,
    2011.

\bibitem{twisted} J. P. Torres and L. Torner, Eds., {\it Twisted Photons: Applications of Light with Orbital Angular Momentum}, Wiley, Hoboken, NJ, USA, 2011.

\bibitem{franke} S. Franke-Arnold, L. Allen, M. Padgett, ``Advances in optical angular momentum'', {\it Laser and Photonics Reviews}, vol. 2, no. 4, pp. 299-313, 2008.

\bibitem{beij} M. W. Beijersbergen, R. Coerwinkel, M. Kristensen, and J. P. Woerdman. ``Helical-wavefront laser beams produced with a spiral phaseplate'',
    \emph{Optics Communications}, vol. 112, no. 5-6, pp. 321-327, 1994.

\bibitem{bazh} V. Yu. Bazhenov, M. V. Vasnetsov, and M. S. Soskin, ``Laser beams with screw dislocations in their wavefronts'', \emph{JETP Letters}, vol. 52,
    no. 8, pp. 429-431, 1990.

\bibitem{barnettallen} S. Barnett, L. Allen, ``Orbital angular momentum and nonparaxial light beams'', {\it Optics Communications}, vol. 110, no. 5-6, pp. 670-678, 1994.

\bibitem{vanenk1} S. J. van Enk, G. Nienhuis, ``Commutation rules and eigenvalues of spin and orbital angular momentum of radiation fields'', {\it Journal of Modern Optics},
    vol. 41, no. 5, pp. 963-977, 1994.

\bibitem{vanenk2} S. J. van Enk, G. Nienhuis, ``Spin and Orbital Angular Momentum of Photons'', {\it Europhysics Letters}, vol. 25, no. 7, pp. 497-501, 1994.

\bibitem{zhao} Y. Zhao, J. S. Edgar, G. D. M. Jeffries, D. McGloin, D. T. Chiu, ``Spin-to-Orbital Angular Momentum Conversion in a Strongly Focused Optical
    Beam'', {\it Physical Review Letters}, vol. 99, no. 7, pp. 073901 (4 pages), 2007.

\bibitem{nieminen} T. A. Nieminen, A. B. Stilgoe, N. R. Hechenberg, H. Rubinstein-Dunlop, ``Angular momentum of a strongly focused Gaussian beam'', {\it Journal of Optics
    A: Pure and Applied Optics}, vol. 10, no. 11, pp. 115005 (6 pages), 2008.

\bibitem{santamoto} E. Santamoto, ``Photon orbital angular momentum: problems and perspectives'', {\it Fortschritte der Physik }, vol. 52, no. 11-12, pp. 1141-1153, 2004.

\bibitem{allen2} L. Allen, M. Padgett, M. Babiker, ``The Orbital Angular Momentum of Light'', \emph{Progress in Optics}, vol. 39, pp. 291�372, 1999.

\bibitem{arfken} G. Arfken and H. Weber, {\it Mathematical Methods for Physicists}, Academic Press, Waltham, MA, USA, 2000.

\bibitem{leach1} J. Leach, M. J. Padgett, S. M. Barnett, S. Franke-Arnold, and J. Courtial, ``Measuring the orbital angular momentum of a single photon'',
    \emph{Physical Review Letters}, vol. 88, no. 25, pp. 257901 (4 pages), 2002.

\bibitem{leach2} J. Leach, J. Courtial, K. Skeldon, S. M. Barnett, S. Franke-Arnold, and M. J. Padgett, ``Interferometric methods to measure orbital and spin, or the total angular momentum of a single photon'', \emph{Physical Review Letters}, vol. 92, no. 1, pp. 013601 (4 pages), 2004.

\bibitem{gao} C. Gao, X. Qi, Y. Liu, J. Xin, L. Wang, ``Sorting and detecting orbital angular momentum states by using a Dove prism embedded Mach�Zehnder interferometer and amplitude gratings'', {\it Optics Communications}, vol. 284, no. 1, pp. 48-51, 2011.

\bibitem{karimi} E. Karimi, B. Piccilillo, E. Nagali, L. Marrucci, E. Santamoto, ``Efficient generation and sorting of orbital angular momentum eigenmodes of
    light by thermally tuned q-plates'', {\it Applied Physical Letters}, vol. 94, no. 23, pp. 231124 (3 pages), 2009.

\bibitem{slussarenko} S. Slussarenko, V. D'Ambrosio, B. Piccirillo, L. Marrucci, E. Santamoto, ``The Polarizing Sagnac Interferometer: a tool for light orbital
    angular momentum sorting and spin-orbit photon processing'', {\it Optics Express}, vol. 18, no. 26, pp. 27205-27216, 2010.

\bibitem{guo} C. S. Guo, S. J. Yue, G. X. Wei, ``Measuring the orbital angular momentum of optical vortices using a multipinhole plate'', {\it Applied Physics
    Letters},
    vol. 94, no. 23, pp. 231104 (3 pages), 2009.

\bibitem{padgett1} M. J. Padgett and L. Allen, ``Orbital angular momentum exchange in cylindrical-lens mode converters'', \emph{Journal of Optics B: Quantum and Semiclassical
    Optics}, vol. 4, no. 2, pp. S17-S19, 2002.

\bibitem{berkhout1} G. C. G. Berkhout, M. P. J. Lavery, J. Courtial, M. W. Beijersbergen, M. J. Padgett, ``Efficient Sorting of Orbital Angular Momentum States
    of Light'', {\it Physical Review Letters}, vol. 105, no. 15, pp. 153601 (4 pages), 2010.

\bibitem{lavery1} M. P. J. Lavery, D. J. Robertson, G. C. G. Berkhout, G. D. Love, M. J. Padgett, J. Courtial, ``Refractive elements for the measurement of the
    orbital angular momentum of a single photon'', {\it Optics Express}, vol. 20, no. 3, pp. 2110-2115, 2012.

\bibitem{mair} A. Mair, A. Vaziri, G. Weihs, A. Zeilinger, ``Entanglement of the orbital angular momentum states of photons'', \emph{Nature }, vol. 412, no.
    6844, pp. 313-316, 2001.

\bibitem{torres2} J. P. Torres, A. Alexandrescu, L. Torner, ``Quantum spiral bandwidth of entangled two-photon states'', \emph{Physical Review A }, vol. 68, no. 5,
    pp. 050301(R) (4 pages), 2003.

\bibitem{ren} X. F. Ren, G. P. Guo, B. Yu, J. Li, G. C. Guo, ``The orbital angular momentum of down-converted photons'', {\it Journal of Optics B: Quantum and Semiclassical Optics},
    vol. 6, no. 4, pp. 243-247, 2004.

\bibitem{barbossa} G. A. Barbossa, ``Wave function for spontaneous parametric down-conversion with orbital angular momentum'', {\it Physical Review A}, vol. 80,
    no. 6, pp. 063833 (12 pages), 2009.

\bibitem{miatto} F. M. Miatto, A. M. Yao, S. M. Barnett, ``Full characterization of the quantum spiral bandwidth of entangled biphotons'', {\it Physical Review A},
    vol. 83, no. 3, pp. 033816 (9 pages), 2011.

\bibitem{gonzalez} N. Gonzalez, G. Molina-Terriza, J. P. Torres, ``How a Dove prism transforms the orbital angular momentum of a light beam'', {\it Optics
    Express},
    vol. 14, no. 20, pp. 9093-9102, 2006.

\bibitem{torner}  L. Torner, J. P. Torres, and S. Carrasco, ``Digital spiral imaging'', {\it Optics Express}, vol. 13, no. 3, pp. 873-881, 2005.

\bibitem{molina}  G. Molina-Terriza, L. Rebane, J. P. Torres, L. Torner, and S. Carrasco, ``Probing canonical geometrical objects by digital spiral
    imaging'', {\it
    Journal of European Optical Society} vol. 2, pp. 07014 (6 pages), 2007.

\bibitem{torres} J. P. Torres, A. Alexandrescu, and L. Torner, ``Quantum spiral bandwidth of entangled two-photon states'', {\it Physical Review A}, vol. 68, no. 5,
    pp. 050301(R) (4 pages), 2003.

\bibitem{fitz} C. A. Fitzpatrick, D. S. Simon, A. V. Sergienko, ``High-Capacity Imaging and Rotationally Insensitive Object Identification with Correlated
    Orbital Angular Momentum States'', {\it International Journal of Quantum Information}, vol. 12, no. 7, pp. 1560013 (16 pages), 2015.

\bibitem{simonspiral} D.S. Simon and A.V. Sergienko, ``Two-Photon Spiral Imaging with Correlated Orbital Angular Momentum States'', \emph{Physical Review A}, vol. 85,
    no. 4, pp. 043825 (8 pages), 2012.

\bibitem{uribe}    N. Uribe-Patarroyo, A.M. Fraine, D.S. Simon, O.M. Minaeva, A.V. Sergienko, ``Object Identification Using Correlated Orbital Angular Momentum
    States'', \emph{Physical Review Letters} vol. 110, no. 4, pp. 043601 (5 pages), 2013.

\bibitem{vasnetsov} M. V. Vasnetsov, J. P. Torres, D. V. Petrov, and L. Torner, ``Observation of the orbital angular momentum spectrum of a
light beam'', {\it Optics Letters}, vol. 28, no. 23, pp. 2285-2287, 2003.

\bibitem{lavery3} M. P. J. Lavery, F. C. Speirits, S. M. Barnett, M. J. Padgett,
``Detection of a spinning object using light�s orbital angular
momentum'', {\it Science}, vol. 341, no. 6145, pp. 537-540, 2013.

\bibitem{rosales} C. Rosales-Guzm�n, N. Hermosa, A. Belmonte, J. P. Torres,
``Experimental detection of transverse particle movement
with structured light'', {\it Scientific Reports}, vol. 3, article ID 2815, 2013.

\bibitem{lavery4} M. P. J. Lavery, S. M. Barnett, F. C. Speirits, M. J. Padgett, ``Observation of the rotational Doppler shift of a white-light,
orbital-angular-momentum-carrying beam backscattered from a rotating
body'', {\it Optica}, vol. 1 no. 1, pp. 1-4, 2014.

\bibitem{padgettphystod} M. Padgett, ``A new twist on the Doppler shift'', {\it Physics Today}, vol. 67, no. 22, pp. 58-59, 2014.

\bibitem{lee2} H. Lee, P. Kok, J. P. Dowling, ``A quantum Rosetta stone for interferometry'', \emph{Journal of Modern Optics}, vol. 49, no. 14-15, pp. 2325-2338, 2002.

\bibitem{bollinger} J. J. Bollinger, W. M. Itano, D. J. Wineland, and D. J. Heinzen, ``Optimal frequency measurements with maximally correlated
    states'', \emph{Physical
    Review A}, vol. 54, no. 6, pp. R4649-R4653, 1996.

\bibitem{bouyer} P. Bouyer and M. A. Kasevich, ``Heisenberg-limited spectroscopy with degenerate Bose-Einstein gases'', \emph{Physical Review A}, vol. 56, no. 2, pp.
    R1083-R1086, 1997.

\bibitem{dowlinggyro} J. P. Dowling, ``Correlated input-port, matter-wave interferometer: Quantum-noise limits to the atom-laser gyroscope'', {\it Physical Review A},
    vol. 57, no. 6, pp. 4736-4746, 1998.

\bibitem{steinberg1} A. M. Steinberg, P. G. Kwiat, and R. Y. Chiao, ``Dispersion cancellation in a measurement of
the single-photon propagation velocity in glass'' \emph{Physical Review Letters}, vol. 68, pp. 2421-2424, 1992.

\bibitem{steinberg2} A. M. Steinberg, P. G. Kwiat, and R.Y. Chiao, ``Dispersion cancellation and high-resolution time measurements in a fourth-order optical
    interferometer'', \emph{Physical Review A}, vol. 45, no. 9, pp. 6659-6665, 1992.

\bibitem{franson} J. D. Franson, ``Nonlocal cancellation of dispersion'', \emph{Physical Review A}, vol. 45, no. 5, pp. 3126-3132, 1992.

\bibitem{erkmen1}  B. I. Erkmen, J. H. Shapiro, ``Phase-conjugate optical coherence tomography'', {\it Physical Review A}, vol. 74, no. 4, pp. 041601 (4 pages),
    2006.

\bibitem{resch2}  K. J. Resch, P. Puvanathasan, J. S. Lunden, M. W. Mitchell, K. Bizheva, K. , ``Classical dispersion-cancellation interferometry', \emph{Optics
    Express}, vol. 15, no. 14, pp. 8797-8804, 2007.

\bibitem{minaeva} O. Minaeva, C. Bonato, B. E. A. Saleh, D. S. Simon, and A. V. Sergienko, ``Odd- and Even-Order Dispersion Cancellation in Quantum
    Interferometry'', {\it Physical Review Letters}, vol. 102,  no. 10, pp. 100504 (4 pages), 2009.

\bibitem{abouraddy} A. F. Abouraddy, M. B. Nasr,  B. E. A. Saleh, A. V. Sergienko, M. C. Teich,
``Quantum-optical coherence tomography with dispersion cancellation'', {\it Physical Review
A}, vol. 65, no. 5, pp. 053817 (6 pages), 2002.

\bibitem{nasr1} M. B. Nasr, B. E. A. Saleh, A. V. Sergienko, M. C. Teich, ``Demonstration of
Dispersion-Canceled Quantum-Optical Coherence Tomography'', {\it Physical Review Letters},
vol. 91, no. 8, pp. 083601 (4 pages), 2003.

\bibitem{nasr2} M. B. Nasr, D. P. Goode, N. Nguyen, G. Rong, L. Yang, B. M. Reinhard,
B. E. A. Saleh, M. C. Teich, ``Quantum optical coherence tomography of a biological sample'', {\it
Optics Commununications} vol. 282, no. 6, pp. 1154-1159, 2009.

\bibitem{bana} K. Banaszek, A. S. Radunsky, I. A. Walmsley, ``Blind dispersion compensation
for optical coherence tomography'', {\it Optics Commununications}, vol. 269, no. 1, pp. 152-155, 2007.

\bibitem{kalten} R. Kaltenbaek, J. Lavoie, D. N. Biggerstaff, K. J. Resch, ``Quantum-inspired
interferometry with chirped laser pulses'', {\it Nature Physics}, vol. 4, no. 11, pp. 864-868, 2008.

\bibitem{legouet} J. Le Gouet, D. Venkatraman, F. N. C. Wong, and J. H. Shapiro, ``Experimental
realization of phase-conjugate optical coherence tomography'', {\it Optics Letters} vol. 35, no. 7,
pp. 1001-1003, 2010.

\bibitem{mazurek} M. D. Mazurek, K. M. Schreiter, R. Prevedel, R. Kaltenbaek, K. J. Resch, ``Dispersion-cancelled biological imaging
with quantum-inspired interferometry'', {\it Scientific Reports}, vol. 3, article ID 1582 (5 pages), 2013.

\bibitem{branning} D. Branning, A. L. Migdall, and A. V. Sergienko, ``Simultaneous measurement of group and phase delay between two photons'', \emph{Physical
    Review A} vol. 62, no. 6, pp. 063808 (12 pages), 2000.

\bibitem{dauler} E. Dauler, G. Jaeger, A. Muller, and A. Migdall, �Tests of a two-photon technique for measuring polarization
mode dispersion with subfemtosecond precision�, \emph{Journal of Research of the National Institute of Standards and Technology}, vol. 104, no. 1, pp. 1-10, 1999.

\bibitem{dangelo} M. D�Angelo, M. Chekhova, Y. Shih,  ``Two-Photon Diffraction and Quantum Lithography'', {\it Physical Review Letters}, vol. 87, no. 1, pp. 013602 (4 pages), 2001.

\bibitem{bjork1} G. Bj\"ork, L. L. S\'anchez-Soto, J. S\"oderholm, ``Entangled-State Lithography: Tailoring Any Pattern with a Single State'', {\it Physical Review Letters}, vol. 86, no. 20, pp. 4516-4519, 2001.

\bibitem{bjork2} G. Bj\"ork, L. L. S\'anchez-Soto, J. S\"oderholm, ``Subwavelength lithography over extended areas'', {\it Physical Review A}, vol. 64, no. 1,
    pp. 013811 (8 pages), 2001.

\bibitem{kothe} C. Kothe, G. Bj\"rk, S. Inoue, M. Bourennane, ``On the efficiency of quantum lithography'',
{\it New Journal of Physical}, vol. 13, pp. 043028 (18 pages), 2011.

\bibitem{dorner} U. Dorner, R. Demkowicz-Dobrzanski, B. J. Smith, J. S. Lundeen, W. Wasilewski, K. Banaszek, and I. A. Walmsley, ``Optimal quantum phase
    estimation'', \emph{Physical Review Letters}, vol. 102, no. 4, pp. 040403 (4 pages), 2009.

\bibitem{demkow} R. Demkowicz-Dobrzanski, U. Dorner, B. J. Smith, J. S. Lundeen, W. Wasilewski, K. Banaszek, and I. A. Walmsley, ``Quantum phase
    estimation
    with lossy interferometers'', \emph{Physical Review A}, vol. 80, no. 1, pp. 013825, 2009.

\bibitem{lee} T. W. Lee, S. D. Huver, H. Lee, L. Kaplan, S. B. McCracken, C. Min, D. B. Uskov, C. F. Wildfeuer, G. Veronis, and J. P. Dowling, ``Optimization of quantum interferometric metrological sensors in the presence of photon loss'', \emph{Physical Review A}, vol. 80,
    no. 6,  pp. 063803, (10 pages), 2009.

\bibitem{shankar} R. Shankar, {\it Principles of Quantum Mechanics, 2nd ed.}, Plenum Press, New York, NY, USA, 2011.

\bibitem{peres} A. Peres, {\it Quantum Theory: Concepts and Methods}, Kluwer Academic Publishers, Dordrecht, Netherlands, 1995.

\bibitem{griffiths} D. J. Griffiths, {\it Intro. to Quantum Mechanics, 2nd ed.}, Pearson Education Ltd., Essex, UK, 2004.

\bibitem{bell1} J. S. Bell, ``On the problem of hidden variables in quantum mechanics'', \emph{Reviews of Modern Physics}, vol. 38, no. 3, pp. 447-452, 1966.

\bibitem{bell2} J. S. Bell, "On the Einstein Podolsky Rosen Paradox", \emph{Physics} vol. 1, no. 3, pp. 195-200, 1964.


\bibitem{aspect} A. Aspect, P. Grangier, G. Roger, ``Experimental Tests of Realistic Local Theories via Bell's Theorem'', \emph{Physical Review Letters},
    vol. 47, no. 7, pp. 460-463,  1981.

\bibitem{edamatsu2} K. Edamatsu, ``Entangled Photons: Generation, Observation, and Characterization'', {\it Japanese Journal of Applied Physics},
vol. 46, no. 11, pp. 7175-7187, 2007.

\end{thebibliography}
\end{document}